# IOLLVM: ENHANCED VERSION OF OLLVM

Chengyang Li   Tianbo Huang, Xiarun Chen,
Chenglin Xie and Weiping Wen

School of Software and Microelectronics, Peking University, Beijing, China

*ABSTRACT*

*Code obfuscation increases the difficulty of understanding programs, improves software security, and, in particular, OLLVM offers the possibility of cross-platform code obfuscation. For OLLVM, we provide enhanced solutions for control flow obfuscation and identifier obfuscation. First, we propose the nested switch obfuscation scheme and the in-degree obfuscation for bogus blocks in the control flow obfuscation. Secondly, the identifier obfuscation scheme is presented in the LLVM layer to fill the gap of OLLVM at this level. Finally, we experimentally verify the enhancement effect of the control flow method and the identifier obfuscation effect and prove that the program's security can be further improved with less overhead, providing higher software security.*

*KEYWORDS*

*Software Protection, Code Obfuscation, Control Flow Obfuscation, Identifier Obfuscation, LLVM.*

## 1. INTRODUCTION

Software protection [1-3] has become more attention-grabbing with the increased awareness of privacy and copyright, especially when the physical and virtual economies are becoming more and more integrated. According to the medium of intervention, existing software protection can be divided into hardware-level [4-6] and software-level [7-9] protection. The former can theoretically provide more robust protection because it can provide a relatively controlled environment. The latter has advantages in cost and applicability compared to the former. Especially with the development of IoT, the security level needed in different scenarios, and the availability of limited computing resources, software-level protection is unavoidable. Considering the broad applicability and objective security obtained for low cost, we focus on software-level protection. Software-level protection can be divided into protection before being damaged and forensics after being damaged, such as protecting core technologies through code obfuscation [10, 11], encryption and decryption [12, 13], software watermarking [7, 14], or software birthmark [15, 16] for effective tracking of software copyright.

Moreover, from the attacker's perspective, it can be divided into anti-static debugging and anti-dynamic debugging. Code obfuscation is effective against static analysis, and virtual machine protection [17, 18] is effective against dynamic analysis. However, virtual machine protection's huge overhead in time and space does not allow for a wide range of landing in the actual production environment. Furthermore, the low overhead of code obfuscation provides considerable security, making code obfuscation continue to occupy an essential role among the many software protection techniques available today. As for code obfuscation, it can be further divided from the hierarchy of roles into source code level [19], intermediate code [20], and binary file [21] obfuscation. Because the intermediate representation is easier to adapt to new front and back ends and easier to carry out optimization work, we propose the code obfuscation scheme on LLVM intermediate representation (LLVM IR).

OLLVM (Obfuscator-LLVM) [20] has implemented a mature and effective obfuscation system on LLVM IR, but currently, OLLVM has some problems as follows. First, OLLVM is based on LLVM 4.0, and as of January 2022, LLVM has released LLVM13. Coupled with the large API gap between different versions of LLVM, OLLVM cannot use some new API features. The adaptation to the newly emerging front-end and back-end may also be problematic. Although different open-source communities [22, 23] have implemented OLLVM migration to new LLVM versions, it is inescapable that existing OLLVM obfuscation techniques are developed based on previous LLVM versions. Furthermore, OLLVM, as a mature obfuscation framework, has had different anti-obfuscation methods implemented for it, and the security that OLLVM itself can provide is questionable. Finally, the existing OLLVM provides obfuscation features including, bogus control flow and flattening at the control flow level and instruction substitution at the instruction level, but does not provide identifier [24] related obfuscation methods. Therefore, this paper implements the iOLLVM system to provide new obfuscation methods at the control flow and identifier levels to address the issues mentioned here.

Following are the main contributions of this paper:
1. The paper proposes enhanced flattening processing and adds in-degree obfuscation for OLLVM control flow obfuscation to resistance to existing scripting attacks;
2. The paper also proposes four algorithms in identifier obfuscation to compensate for the lack of OLLVM.

The rest of the paper is organized as follows: Section 2 introduces the overview of the iOLLVM. Section 3 describes the details of identifier algorithms, and the experiment results are presented in section 4. Finally, in section 5, we end up with a conclusion and a description of future work.

## 2. THE COMPONENTS OF IOLLVM

Considering the stability of the new version of LLVM API, the prototype iOLLVM obfuscation system was implemented on LLVM 10, giving effect enhancements at the control flow obfuscation level and the identifier obfuscation level, respectively. The general framework diagram is shown in Figure 1.

Based on the existing OLLVM, iOLLVM proposes an enhanced obfuscation implementation at the control flow level and complements the obfuscation method at the identifier level. The overall flow of the obfuscation system is as follows: first, the source code is transformed into LLVM IR code by the front-end tools provided by LLVM, such as clang, flang, denoted as xclang in Figure 1. Then, the LLVM IR code is obfuscated by the new modules provided by iOLLVM: the control flow module and the identifier module. They can be called and used separately or nested. Finally, the obfuscated IR files generate platform or environment-ready executables by specific back-end programs.

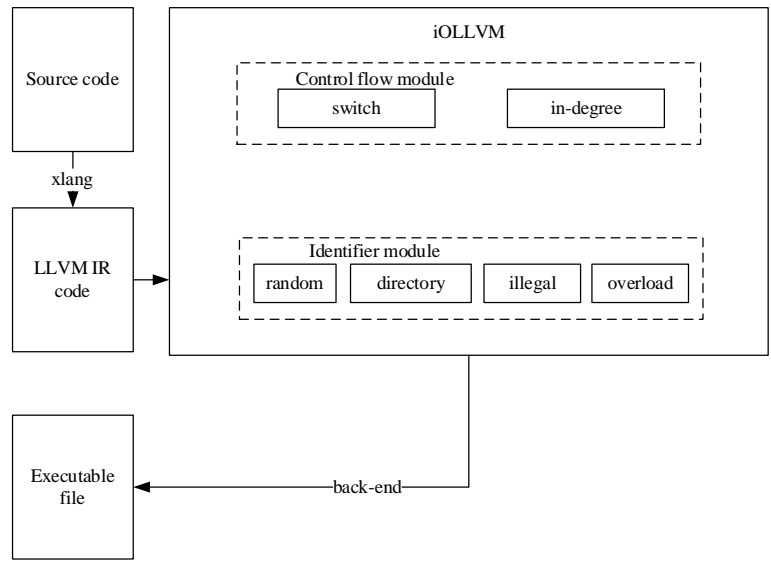

Figure 1. iOLLVM general framework diagram

## 2.1. Control Flow Obfuscation

According to Collberg's classification of code obfuscation [10], control flow obfuscation intends to effectively hide the program's control flow and make the attacker's path analysis more difficult. OLLVM provides control flow obfuscation, including bogus control flow and flattening processing [20]. However, using some IDA Python script files [25] to target the features of OLLVM obfuscation processing can effectively remove the added flattened obfuscated code. In the meantime, if a cracker can infer that control flow obfuscation has been performed based on the control flow graph processed by the automated tool, this itself is also valuable information for further analysis of the program. Moreover, at the same time, the bogus control flow design currently provided does not take into account the attack of the degree of entry analysis. Therefore, this paper proposes a counter-common sense nested switch flattening process and an in-degree obfuscation to prevent incidence analysis.

### 2.1.1. Nested Switch

The original purpose of flattening was to disrupt the program's execution flow. In LLVM IR, the basic blocks that initially had an upper and lower hierarchy are transformed into the same hierarchical relationship. Considered the convenience and the cross-platform nature of the program, the general implementation uses the dispatcher architecture of the switch, and the processed program has a prominent feature while bringing security, namely the presence of the loop structure. The switch structure distributes the basic blocks. From the point of view of hiding the flattened processing and resisting script analysis, the switch structure is created again in the flattened generated switch structure, and a new switch is created in each basic block corresponding to the case.

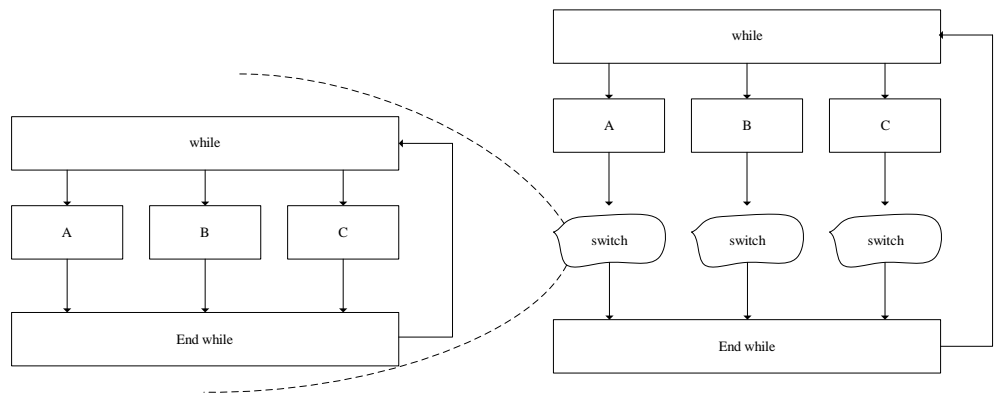

Figure 2. Nested switch structure

As shown in Figure 2, the part in the imaginary coil is noted as procedure F: the general flattening operation. The new switch structure is created inside the basic block, noted as procedure N. The procedure for adding boolean or numerical operations against variable v is noted as T. The steps of the nested switch are as follows.

Step 1: Perform procedure F on input program (or LLVM IR file) P0 to realize the flattening operation; the obfuscated program is noted as P1.
Step 2: Iterate through the basic blocks in P1 at the same level, and record them as the setBlocks, and for each basic block inside, process N to achieve the nesting of switch structure, and the obfuscated program is recorded as P2.
Step 3: For the switch generated at the P1 level, the jumping variable is recorded as the case-outer; for the switch generated at the P2 level, the jumping variable is recorded as the case-inner. The procedure T is added to the basic block corresponding to the case-inner, but only one block is left unprocessed to ensure normal program execution. The obfuscated program is denoted as P3.

P3 is the program after nested switch processing. The process of creating the nested switch structure requires attention to the creation of the basic block (or called the bogus block) and the setting of the jumping variables (case-outer, case-inner) used by the internal and external switches for distribution.

1. Because it involves the construction of a new switch structure, corresponding to the creation of new basic blocks, to enhance the similarity with the native program, drawing on the idea of bogus control flow, based on the program's native basic blocks to transform. Further, the number of bogus blocks is created because different space overhead scenarios are expected to be different. The default value is the number of cases of the outer switch while providing external parameters specified at the time of use.
2. Because the internal and external switch structure is involved, to strengthen the confusion of the internal switch structure and ensure the semantics of the program, the external case values are processed in the internal switch structure. However, there will always be a basic block in the internal randomly generated case that is not processed for the external case. All other basic blocks corresponding to the same level of cases add random boolean or arithmetic operations. This results in a combination of obfuscation processing on data dependencies.

**2.1.2. In-Degree obfuscation**

For the native bogus control flow, it is possible to protect the program by copying and transforming the native basic blocks to construct new bogus blocks embedded in the program through opaque predicates [26]. Generally, in practice, to reduce the space overhead, the obfuscation algorithm only protects against the critical code in the program. So if the attacker can recognize that the bogus control flow protects the program, there is a risk of identifying the bogus blocks. Plus, to hide the difference between the real block and the bogus block, a path to the real block can generally be constructed in the bogus block, so there will be: in general, the real block in-degree will be greater than that of the bogus block. Moreover, this can play a more significant role in excluding bogus blocks. Therefore, to resist the attacker's degree of entry analysis, the degree of entry of the bogus block is obfuscated.

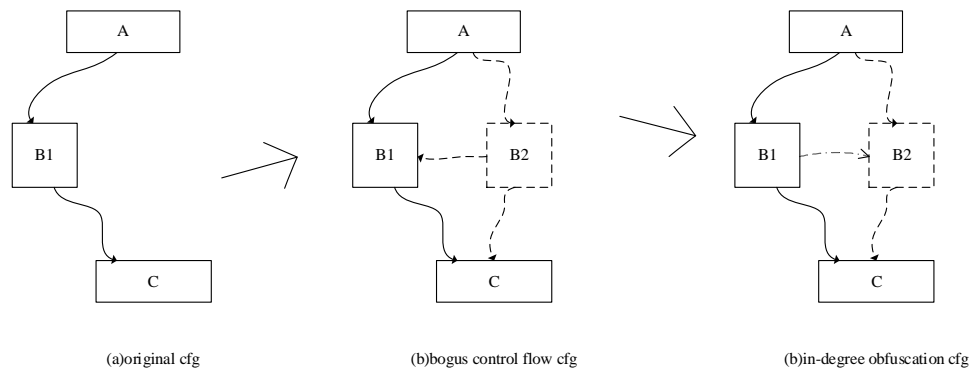

(a)original cfg        (b)bogus control flow cfg        (b)in-degree obfuscation cfg

Figure 3. In-degree obfuscation

Figure 3(a) shows the sample program with a direct jumping relationship between the basic blocks A, B1, and C. Figure 3(b) shows the schematic diagram after adding the bogus control flow processing. Figure 3(c) shows the schematic diagram after adding the in-degree obfuscation processing, noting that B1 points to the edge of B2.

The fundamental point of entry analysis is to make bogus blocks have more entries than the real block by opaque predicates, i.e., adding opaque predicates to the real block to realize the addition of false jump relations. However, suppose there are already jump relations in the original real block without splitting the real block. In that case, the logic can be combined with the idea of flattening, transforming different branches into switch-case jump forms, and adding opaque predicates to the internal cases.

**2.2. Identifier Obfuscation**

During the reverse analysis of the program, the strings and identifiers exposed by the program will play a significant role in guiding the attacker to the specific location in the program based on the strings. They may even expose some core algorithm design and other information about the program. Therefore, it is necessary to hide the string and identifier information in the program effectively. Although OLLVM does not provide obfuscation protection for strings, both Armariris [27] and Hikari [28] give specific open-source schemes to protect certain programs. However, no explicit protection method is given for identifiers. Although the identifier information in the program can be partially removed using strip against the program, it is beneficial but not harmful to provide the obfuscation of identifiers in terms of the completeness of security and the uncertainty of the usage scenario. Therefore, this paper adds to this aspect by providing four obfuscation algorithms for identifiers in LLVM IR: Random, Directory, Illegal, and Overload.

## 3. ALGORITHMS FOR IDENTIFIER OBFUSCATION

This paper proposed four algorithms to provide software security, namely, Random, Directory, Illegal, and Overload. The first three belong to the general category of substitution algorithms, while the fourth is a separate algorithm category.

| Algorithm 1 Substitution algorithm |
| --- |
| 1: init parameters that algorithm needs |
| 2: Read LLVM IR file |
| 3: get all custom identifiers noted as setNames |
| 4: select a algorithm from Set(Random, Directory, Illegal) |
| 5: for item in setNames: |
| 6:     change item in global |
| 7: end |

Among them, the replacement data sources for Directory and Illegal need to be initialized beforehand, and the replacement data sources for Random can be generated beforehand or dynamically at runtime.

| Algorithm 2 Overload algorithm |
| --- |
| 1: Read LLVM IR file |
| 2: get all custom identifiers noted as setNames |
| 3: for item in setNames: |
| 4:     generate a new function name item with random parameters, noted as newname |
| 5:     if(newname is legal) |
| 6:         added it to LLVM module |
| 7: end |

The critical point of the overload algorithm is that some languages support the overloading feature internally. When designing, we need to judge whether the overloading feature is satisfied based on the name mangling of the parameter list setting information. Next, the algorithm details are explained as follows.

Random algorithm: First, obtain the identifiers in the program. Randomly generate 11 identifiers containing letters, numbers, and special symbols and conform to the naming convention. Then, perform the global replacement for each identifier.

Directory algorithm: Unlike the idea of randomly generating identifiers for replacement, the purpose of the obfuscation dictionary is to use standard identifiers to replace identifiers in the program. Standard identifiers can play a preemptive role in obfuscation, unlike meaningless random characters.

Illegal algorithm: There will be reserved words in the program design, and it is not possible to use reserved words to name the identifiers, but since all characters are coded at the code set, characters of different codes set may be highly similar in appearance, so the Greek letters are used to replace the English letters thus playing the role of replacement confusion.

Overload algorithm: To further enhance the obfuscation of the program, the program is processed using the overload idea. Also, to avoid the name-mangling mechanism of the program itself, the parameter list information is kept when obtaining the identifier information. Then the overloaded identifier of the virtual parameter list is constructed.

Because the idea of the first three is the idea of substitution, the fourth is the idea of addition. Therefore, in the user use, provide the default way, from the first three randomly choose one, and then with the fourth to reuse the previous confusing characters as much as possible.

## 4. EXPERIMENTS

We carry out three experiments: a) control flows obfuscation versus OLLVM; b) identifier obfuscation effects; c) time and space overhead.

Experiment environment: Ubuntu18.04; Inter(R) Core(TM) i5-4210U; 4G Memory; LLVM10.

Experimental data is selected from the algorithm library [29] implemented in c language on GitHub. In order to facilitate the automatic execution of the script, the part of the algorithm that involves input is converted from user input to fixed-value input. Finally, only 123 source files are kept.

### 4.1. Experiment 1- Control Flows Obfuscation Versus OLLVM

Function kthSmallest in kth_smallest.cpp in the test set is used to illustrate the effect of obfuscation. First, use clang to process the source program to get the corresponding LLVM intermediate code bc file; then use the so file generated by the obfuscation system to obfuscate the specified file. Here we can further specify the function to be obfuscated. The sample below specifies the function kthSmallest for processing. Because C++ has a name mangling mechanism compiled into intermediate code, so the function name seen in the intermediate code layer is _Z11kthSmallestP8TreeNodei. But the user only needs to provide the function's name in the source code when specifying it. Finally, Graphviz is used to generate the graph summary as follows.

### 4.1.1. Nested Switch

The control flow diagram of the function before obfuscation is shown in Figure 4, the control flow diagram after flattening with OLLVM is shown in Figure 5, and the control flow diagram after nest switch processing is shown in Figure 6. For ease of display, only show the logical relationships of the basic blocks while the instructions in Figure 5 and Figure 6 are removed.

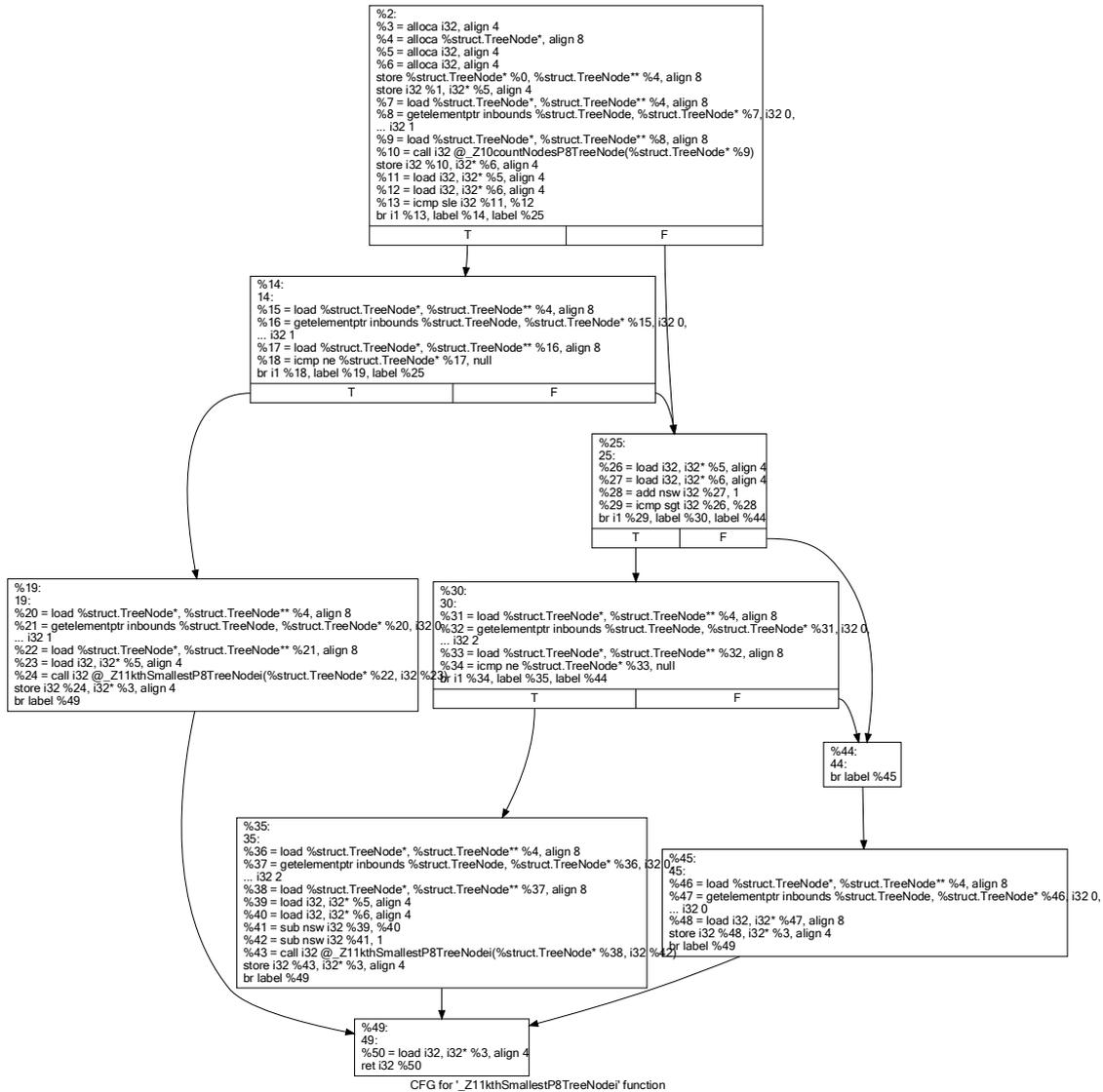

Figure 4. KthSmalles: origin cfg

As shown in Figure 5, the original basic block relationship between the upper and lower levels is transformed into the same hierarchy level, with efficient scheduling through the switch. The attacker can infer flattened processing based on the current control flow graph and then use the existing attack script to approximate the program.

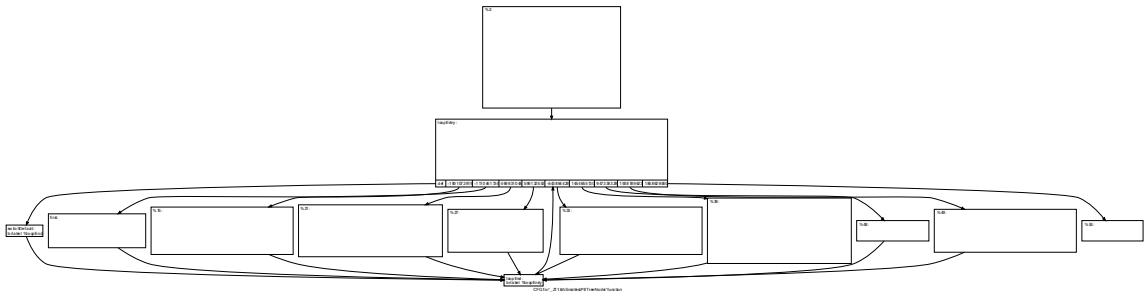

Figure 5. KthSmalles: ollvm flattening cfg

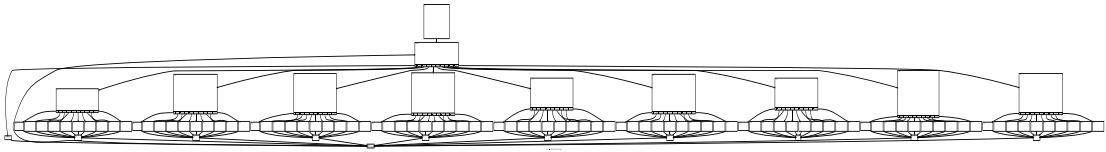

Figure 6. KthSmalles: nested switch flattening cfg

As shown in Figures 4, 5, and 6, nested switches can make an order of magnitude increase in complexity. The nested structure is different from the existing flattening processing. Even if an attacker uses an existing attack script for processing, it can effectively resist the existing attack script because the characteristics of the obfuscation algorithm are different. In addition, for control flow obfuscation, the similarity of basic blocks (BB), jump instructions(JI), and functions(F) after obfuscation and the similarity of programs(P) before and after obfuscation can be used as indicators to measure the effectiveness of obfuscation. Therefore, BinDiff [30] counts the above four indicators in the test set. Table 1 shows the statistics of the metrics after OLLVM obfuscation, and Table 2 shows the statistics of the metrics after obfuscation using nested switches and in degrees.

Table 1. Similarity of OLLVM obfuscated files to source files

| OLLVM flatten | Average value | Minimum value | Maximum value | Standard deviation |
|---|---|---|---|---|
| similarity of BB | 68.33 | 40.2 | 100 | 12.93 |
| similarity of JI | 36.55 | 10.9 | 100 | 21.4 |
| similarity of F | 76.72 | 55 | 92.6 | 8.79 |
| similarity of P | 0.74 | 0.56 | 0.99 | 0.1 |

Table 2. Similarity of nested switch, in-degree obfuscation files to source files

| Nested switch, in-degree | Average value | Minimum value | Maximum value | Standard deviation |
|---|---|---|---|---|
| similarity of BB | 29.09 | 6.7 | 100 | 21.01 |
| similarity of JI | 14.19 | 1.6 | 100 | 21 |
| similarity of F | 76.67 | 55 | 92.6 | 8.79 |
| similarity of P | 0.5 | 0.34 | 0.99 | 0.15 |

In terms of jump instructions, methods in this paper are on average 61% lower than OLLVM and 57% lower in terms of similarity of basic blocks. The reason is twofold: firstly, this method constructs more basic blocks and jump instructions, which significantly reduces the similarity with the native program and increases the confusion; secondly, the jump instructions in some of the original basic blocks are changed, so the decrease in the similarity of jump instructions is more prominent.

There is essentially no difference in the number of functions considering the effect of errors generated by BinDiff decompilation. The similarity between the obfuscated program and the original program, the method in this paper decreases by 32% compared to OLVM. Meanwhile, the maximum values of the four metrics, OLLVM, and this paper's method agree, i.e., for some particular files, the effect of the control flow obfuscation method proposed in this paper is consistent with OLLVM. However, the minimum values and standard deviations show that this paper's method performs better than OLLVM on the applicable files. It should be added that the similarity is one of the indicators of the obfuscation effect. However, suppose we only increase the number of basic blocks and jump instructions. In that case, such processing will not play a corresponding resistance to the substantial reverse cracking. However, this paper combines the construction of basic blocks and jump instructions with the control flow and data flow of the original program itself, so the obfuscation effect is guaranteed to a certain extent.

### 4.1.2. Indirect Jump

The control flow diagram of the function before obfuscation is shown in Figure 4, and the function after obfuscation using the proposed obfuscation process for in-degree analysis is shown in the following figure, Figure 7. The image has been simplified for easy display.

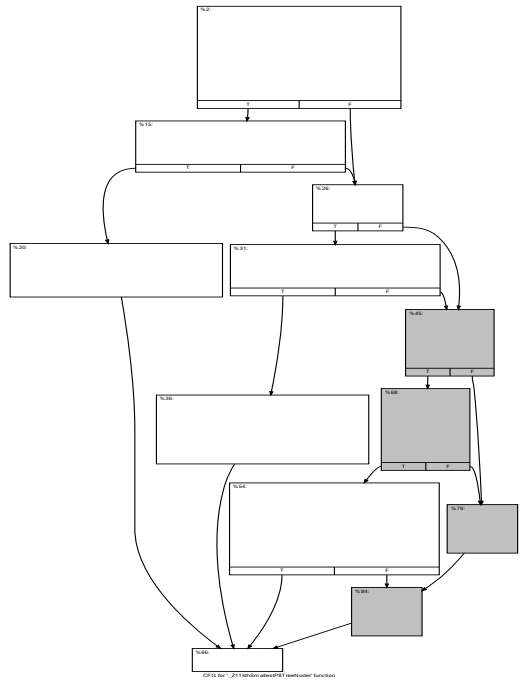

Figure 7. In-degree obfuscation effect

The grey blocks in Figure 7 are newly bogus, while white blocks are original blocks in the origin program. It can be seen that by introducing opaque predicates, the degree of entry of the bogus block is significantly higher than that of the actual block while ensuring the original semantics of the program, thus avoiding the threat of entry analysis. Limited to the constructional characteristics of the algorithm itself, it will have higher security when used simultaneously with other control flow methods.

### 4.2. Experiment 2- Identifier Obfuscation Effects

After using the identifier obfuscation algorithm for the 123 files in the test set, the function identifier replacement rate is 65.2% (4875/7515). In order to ensure the standard semantics of the program and avoid modifications to third-party library functions, only user-defined function names are selected here. It should be noted that for large projects in terms of identifier replacement, we can consider using wllvm [31] to link the project code into a single bc file and then obfuscate the identifiers for the bc file to avoid errors caused by symbolic links.

### 4.3. Experiment 3-File Size and Performance Penalty of Protected Program

Table 3. Time and space overhead of obfuscation methods

| Obfuscation Method | Time overhead | Space overhead |
| --- | --- | --- |
| OLLVM Flattening | 1.01 | 1.05 |
| Identifier obfuscation | 0.98 | 1.00 |
| Nested switch, in-degree | 1.02 | 1.59 |

The space overhead is measured by comparing the increase of the file size after obfuscation with the increase in file size before obfuscation. As for time overhead, we use a script to count the time of multiple file executions before and after obfuscation. Furthermore, taking the average value as a benchmark for comparison. From Table 3, we can see that the control flow enhancement scheme proposed in this paper has a slight increase in time and space compared to OLLVM. In particular, considering that the program may run slightly differently in time under different execution states, the existing test sets are all small volume algorithm files, so they are more sensitive to execution time. Therefore, it can be considered that the identifier obfuscation has almost no overhead impact in terms of time and space.

It should also be noted that in large projects, as the volume of the protected code increases, the corresponding time and space overhead also increases, and the two should be linear. When using, different parameter values should be set in conjunction with specific usage scenarios to meet the security and performance requirements of the scenario.

## 5. CONCLUSIONS

This paper addresses the lack of strength of OLLVM obfuscation in control flow protection and the gap in identifier obfuscation by proposing two broad categories of enhancements. In control flow obfuscation, first, adding nested switches at the control flow level and adding the switch structure again in the flattened code, thus increasing the complexity of the code while resisting existing scripting attacks; second, proposing an in-degree treatment for bogus blocks to increase the confusion of bogus blocks further. Further, at the level of identifier obfuscation, four algorithms are proposed and bridge the gap of OLLVM in identifier obfuscation. By comparing with OLLVM, this paper can significantly improve the original control flow complexity in obfuscation effect; replace 65.2% of custom identifiers while guaranteeing program functionality. Furthermore, the time overhead from obfuscation is almost negligible. The space overhead is at 1.5 times.

In future work, we will pay attention to generating more secure opaque predicates and are not limited to the number-theoretic model. Meanwhile, the practical effectiveness of existing obfuscation algorithms in large projects remains tested. Therefore, we will focus on how to provide more accessible use of the obfuscation framework model in large projects.

**ACKNOWLEDGEMENTS**

Thanks to every author who participated in the paper, it is a joint effort to produce this article.

**AUTHOR**

**Chengyang Li**, born in 1996, M. S. His research interests include code obfuscation.

**HUANG Tianbo**, born in 1997, M. S. His research interests include cyberspace security, malicious code detection and code obfuscation.

**Xiarun Chen**, born in 1997, M.S. His research interests include system and network security, blockchain security and malicious code detection.

**ChenglinXie**, born in 1996, M. S. His research interests include Operating System Security, Intrusion Detection and Malware Detection.

**WEN Weiping**, born in 1976, Ph. D. professor. His research interests include system and network security, big data and cloud security, intelligent computing security.